\documentclass{Interspeech}

\interspeechcameraready

\title{Can We Trust Machine Learning? The Reliability of Features from Open-Source Speech Analysis Tools for Speech Modeling}

\author[affiliation={1}]{Tahiya}{Chowdhury}
\author[affiliation={2}]{Veronica}{Romero}

\affiliation{Department of Computer Science}{Colby College}{USA}
\affiliation{Department of Psychology}{Colby College}{USA}
\email{tahiya.chowdhury@colby.edu, veronica.romero@colby.edu}

\keywords{autism, reliable machine learning, speech processing tools.}

\newcommand{\blue}[1]{\textcolor{black}{#1}}
\usepackage{comment}
\usepackage[table]{xcolor}
\definecolor{mycolor}{RGB}{220,220,220} 

\begin{document}

\maketitle

\begin{abstract}
    
Machine learning-based behavioral models rely on features extracted from audio-visual recordings. The recordings are processed using open-source tools to extract speech features for classification models. These tools often lack validation to ensure reliability in capturing behaviorally relevant information. This gap raises concerns about reproducibility and fairness across diverse populations and contexts. Speech processing tools, when used outside of their design context, can fail to capture behavioral variations equitably and can then contribute to bias. We evaluate speech features extracted from two widely used speech analysis tools, OpenSMILE and Praat, to assess their reliability when considering adolescents with autism. We observed considerable variation in features across tools, which influenced model performance across context and demographic groups. We encourage domain-relevant verification to enhance the reliability of machine learning models in clinical applications.
\end{abstract}

\section{Introduction}

Clinical conditions that manifest neurological, cognitive, and behavioral symptoms are typically diagnosed with standardized tools. Consider Autism Spectrum Disorder, which refers to a set of developmental conditions that restrict an individual's ability for social communication and interaction, among other things. Currently, the clinical measures that characterize symptoms of autism are assessed by diagnostic instruments, which depend on subjective ratings of behavioral symptoms by an expert. While these measures are clinically valid and well understood, the current assessment process is cognitively demanding and costly. Many recent studies have shown promising results through the use of automated machine learning-based approaches using speech and language features as behavioral measures to classify clinical conditions such as Autism~\cite{rybner_vocal_2022, chowdhury23_interspeech, chi2022classifying, briend2023voice} and Dementia~\cite{steinert2021audio, melistas23_interspeech}. This approach relies on open-source software tools to extract behavioral features from speech samples recorded from individuals.

While automated diagnostic tools have the potential to support existing diagnostic processes by providing objective behavioral measures, several challenges remain to translate them into clinical practice. First, the widely available open-source tools provide low-level speech features extracted from audio recordings, which are then used to model behavior and predict a diagnostic label of an individual. However, these features do not typically go through any process to check their validity in a given clinical context to be contextualized for behavioral modeling~\cite{stegmann2020repeatability, berisha22_interspeech}. Second, various open-source tools exist to conveniently extract a large number of speech features including OpenSMILE~\cite{10.1145/1873951.1874246, kim2021analyzing, yang23u_interspeech}, Praat~\cite{stegmann2020repeatability,briend2023voice}, and ProsodyPro~\cite{chen24b_interspeech}, to name a few. However, these tools are designed for more general-purpose speech processing applications whose default settings may not translate well in clinical applications. Many design choices are involved in the feature extraction as well as the modeling process that is guided by the intention to improve classification scores instead of domain-specific knowledge~\cite{fusaroli2017voice}. As a result of this lack of standardization, over-optimistic results are reported that are actually the consequence of issues such as data leakage (e.g., sampling bias)~\cite{kapoor2023leakage}, model overfitting on small sample size~\cite{berisha22_interspeech}, and lack of generalizability across different contexts~\cite{rybner_vocal_2022}. Finally, in the absence of clinically and contextually relevant features to characterize clinical conditions that work for diverse demographic groups and languages~\cite{fusaroli2022toward}, models tend to generalize poorly for marginalized groups in real-world clinical settings.

In this work, we evaluate behavioral features measured using commonly used open-source speech analysis tools using audio data collected for autism diagnosis of adolescents. Specifically, we look at OpenSMILE~\cite{10.1145/1873951.1874246} and Praat~\cite{boersma2007praat} as they provide speech features directly using default settings without any pre-processing, making them more accessible for model development. We focused on pitch and speech rate~\cite{ochi2019quantification, chen24b_interspeech, kiss2012quantitative, bone2015acoustic, 10.1016/j.csl.2015.09.003} based on prior literature where speech has been used as a biomarker to classify diagnostic label of autism. Specifically, Pitch standard deviation (pitch variability), Pitch range (pitch modulation during a specific conversation context)~\cite{bonneh2011abnormal} and Loudness~\cite{kim2021analyzing} are used to represent a person's ability to adapt in a conversational setting. High mean pitch is generally used as a marker for distinctive speech patterns for ASD groups. Speech rate is generally used to understand the speed at which an individual speaks spontaneously, and slow speech rate can distinguish children (at developmental age) diagnosed with autism from their neurotypical peers~\cite{flipsen2006syllables}. We compare these five features measured with the two tools to understand the reliability of the measures and their impact on machine learning-based classification performance. We aim to answer the following questions in this work:

\begin{itemize}
    \item Are some speech features provided by widely used speech processing tools (OpenSMILE and Praat) measured the same way?
    \item Do these tools measure speech features differently based on the gender of participants' data?
    \item Do these tools measure speech features differently based on diagnostic group membership?
    \item Do machine learning classification models perform robustly for each diagnostic group regardless of the speech processing tool used?

\end{itemize}

\section{Data}

We utilized data collected during sessions of the Autism Diagnostic Observation Schedule - Second Edition (ADOS-2), a standardized assessment tool designed to evaluate ASD-related impairments~\cite{lord_autism_2000}. In this evaluation, an adolescent (`Participant') interacts with a certified adult assessor through a series of semi-structured activities that assess the child's behavioral patterns. The dataset for this study comprised 14 distinct `Tasks' from Module 3 of the ADOS-2, which is specifically designed for verbally fluent children and adolescents. Depending on the task, the participant may be asked to narrate a story, play with toys, act out a cartoon, or respond to questions on topics such as emotions, loneliness, and friendship. Since our focus is on features learned across different contexts, we incorporated all 14 tasks from the module. 

Our dataset consisted of recordings from 29 adolescents (14 with ASD and 15 typically developing, TD) between the ages of 10 and 15 (\emph{M}: 12.27, \emph{SD}: 1.75). There were 21 male and 8 female participants. Each ADOS-2 session lasted approximately 40–60 minutes and was administered by the same psychologist. For this study, we first generated individual recordings for each subtask using annotations made by a research assistant based on video recordings of the sessions. The average length of these video segments is about five minutes per task. Speaker diarization was performed using \emph{pyannote}~\cite{Bredin2020}, and \emph{PyDub}~\cite{pydub} was employed to segment the audio further based on speaker identity and turn boundaries obtained from \emph{pyannote}. This process resulted in utterance-level audio files for both the participant and the psychologist. For the current project, we used the participants' audio files for speech feature extraction. After filtering out empty utterances, our final dataset contained an average of 116 utterances per participant. 

\section{Method}
We first extracted speech features from the audio recordings of the participants using the two speech processing tools considered here: \emph{OpenSMILE}~\cite{10.1145/1873951.1874246} and \emph{Praat}~\cite{boersma2007praat} to compare them across demographic groups and their performance in machine learning classification models. 

\subsection{Features}

For OpenSMILE, we used the Geneva Minimalistic Acoustic Parameter Set (eGeMAPSv02) for Voice Research and Affective Computing~\cite{7160715} as prior work has found these features useful for autism and conversation settings~\cite{stegmann2020repeatability}. For pitch, we used fundamental frequency (F0), 
from which we calculated Mean Pitch, Pitch standard deviation, Pitch range, and Loudness. For speech rate, we used voiced segment per second to estimate words per minute (assuming 1.5 words per segment). This estimated word per minute is converted to syllables per second, which we used as the measure of speech rate~\cite{flipsen2006syllables}.

For Praat, we used Parselmouth~\cite{parselmouth}, a Python wrapper for Praat for ease of use. Praat calculated the pitch for each audio file, from which we calculated the mean, standard deviation, range, and loudness using intensity during the audio. For each utterance audio, we used intensity to detect syllables and count them. The voiced audio duration was used to calculate the syllables per second, the measure used for the speech rate feature.

Using the utterance level feature sets calculated above, we further computed task-level aggregate features for each task per child. This provided us with the mean value for each feature per child by both tools, which we used to compare the features from each of the tools to each other.

\subsection{Statistical Analyses}

To answer whether OpenSMILE and Praat measure the five features differently, we used the task-level aggregate feature calculated using both tools and proceeded to extract the mean difference for each participant per task. We then calculated the mean and standard deviation of these differences for each of the five features (Mean pitch, Pitch Std, Pitch range, Loudness, and Speech rate) to compare the feature values measured by each tool. We also performed paired-sample t-tests for each feature pair to assess if any present differences were statistically significant. 

To explore whether Task-level features measured by these tools are different by demographic group membership (Diagnostic group and gender), we performed separate Two-way ANOVA analyses: a set of 2(gender: male or female) x 2(tool: OpenSMILE or Praat); and a second set of 2(diagnostic group: ASD or TD) x 2(tool: OpenSMILE or Praat) to test main effects and possible interaction effects. For cases where a significant interaction effect was found, we performed post-hoc pairwise comparisons with Bonferroni adjustments~\cite{10.1111/j.2517-6161.1995.tb02031.x} to avoid Type I error from performing multiple tests so that we can identify whether an individual tool provided significantly different features based on demographic characteristics (ASD and TD, male and female).

\subsection{Classification}

Machine learning classifiers are typically trained using features extracted from a training set, and then performance is reported on a test set that was unseen during training. Since we are particularly interested in how features measured by each tool perform in distinguishing diagnostic status (ASD and TD), we trained separate random forest classifiers using features from each tool. The task here was to classify children based on diagnostic groups (ASD and TD). 

To ensure generalization for out-of-sample testing, we performed cross-validation by using a leave-$one$-user-out method\footnote{We chose this over commonly used \emph{k}-fold cross-validation to avoid data leakage and ensure our training set did not include information from a child who was also present in our test set.}. We report results averaged over all runs, where one child was put into the test set, and the remaining children were the train set to partly simulate a real-world diagnosis scenario for new participants. All classification results are reported in accuracy, precision, recall, and F1-score. All experiments and models were implemented using the \textit{scikit-learn}, \textit{sciPy}, and \textit{statsmodels} libraries with default parameter settings.

\begin{table}[t]
    \centering
    \caption{Comparison of difference in speech feature values extracted form Praat and OpenSMILE.}
    \begin{tabular}{lcc}
        \toprule
        \textbf{Feature} & \textbf{Mean Diff.} & \textbf{Standard Dev.} \\
        \midrule
        Mean Pitch & 156.9401 & 5.9803 \\
        Pitch Std & 31.8441 & 7.3489 \\
        Pitch Range & 125.6173 & 22.1054 \\
        Loudness  & 4.4793 & 1.0347 \\
        Speech Rate 
        & 48.3723 & 0.8405 \\
  
        \bottomrule
    \end{tabular}
    
    \label{tab:comparison}
\end{table}

\section{Results}

\subsection{Are features measured by OpenSMILE and Praat the same way?}
To answer whether the two tools under consideration, OpenSMILE and Praat, provide the same measure of each feature, we calculated the difference in features measured by the two tools for each participant by task. Since the difference in feature values for various tasks was negligible, we calculated the mean value of these differences and the standard deviation to show the extent of their variability. We report these results in Table~\ref{tab:comparison}.

\begin{figure}[t]
  \centering
  \includegraphics[width=0.80\linewidth]{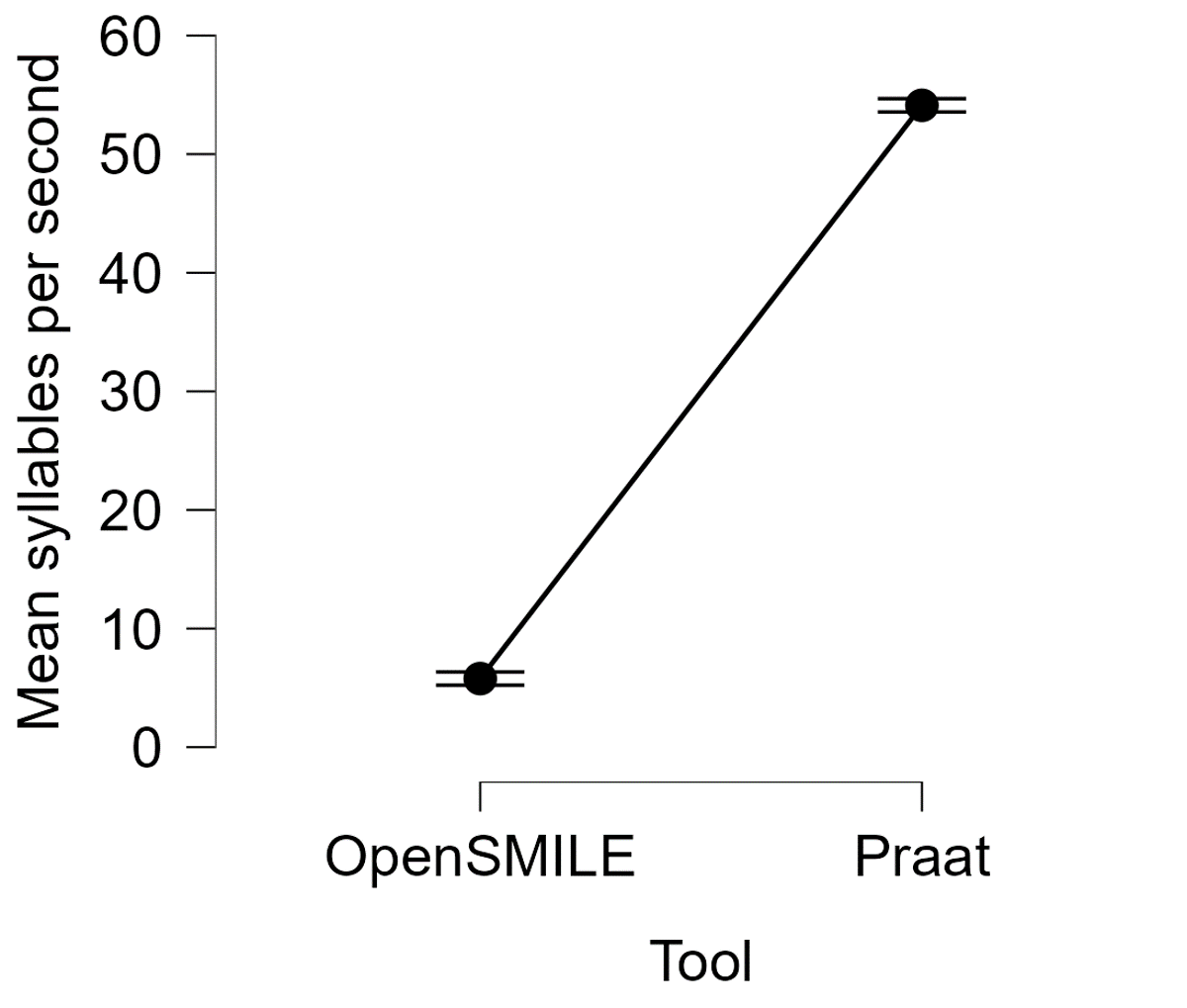}
  \caption{Mean speech rate measured by OpenSMILE and Praat. Note that average adult speech rate is between 4 and 6 syllables/second. }
  \label{fig:speech_production}
  \vspace{-10pt}
\end{figure}

We found each speech feature measured by OpenSMILE and Praat to be significantly different from each other \blue{($p < 0.001$ in Paired sample t-test)}. This would indicate that each model is measuring what we assume to be the same feature differently. While judging the validity of each tool is beyond the scope of the current project, we would like to point out that in considering speech rate, Praat sometimes measures this feature in a way that is behaviorally impossible for the participants to produce. For example, if you consider the results presented in Figure~\ref{fig:speech_production}, OpenSMILE measures an average of 5.78 syllables per second, while Praat calculates a mean of 54.12 syllables per second, which is just not possible for even adults to produce. \blue{Note that diarization noise and segmentation errors can lead to silence, clipped audio, or misattributed turns, affecting speech rate estimates. Praat, which relies on intensity-based syllable detection and voiced duration, is more sensitive to such artifacts. In contrast, OpenSMILE uses heuristic mappings from voiced segments to syllables, offering more stability. Thus, Praat’s higher variability here may reflect artifact sensitivity rather than superior speech modeling}.

\subsection{Do these tools measure speech features differently based on demographic group?}


We performed ANOVA tests to better understand the effects accounted for by the tool used, demographic group membership, and their interaction for each task. We report significant main effects or interactions for a measured feature in a specific task in Table~\ref{tab:anova_combined_results}.

\textbf{Gender Group.} Multiple tasks involved features where we see significant differences between the measure from OpenSMILE and Praat and also between demographic groups. In particular, when considering mean Pitch during `Description' or `Demonstration', the measures provided for male participants were significantly different than those provided for female participants. Importantly, this difference was only present when using Praat. Similarly, calculated Pitch variability (Pitch range), Pitch std, and Speech rate were significantly different based on gender. However, none of these features differed significantly by gender when using OpenSMILE, only when using Praat ($p < 0.05$).

\textbf{Diagnostic Group.} Speech rate during the `Cartoon' task was found to be significantly different when comparing ASD and TD groups ($p<0.01$). However, this difference was only present when using Praat ($p=0.2801$). Similarly, during `Construction', only Praat found a significant effect of the diagnostic group on Pitch std. In general, the results suggest that Praat provides more sensitive calculations when considering possible distinctions present in speech features based on demographic groups compared to OpenSMILE.

\subsection{Do classification results vary based on the tools used?}

To investigate how tools and features extracted by them influence model performance in classifying diagnostic labels, we compared the performance for both tools by the two diagnostic groups separately for each of the 14 tasks (in Table~\ref{tab:classification_tasks_child}).


When considering precision, a measure for the percentage of samples that were classified to the correct label, OpenSMILE and Praat performed similarly. However, the ASD group was misclassified more often when using Praat for `Demonstration' ($0.77$) compared to the TD group ($0.93$). For recall, which measures the percentage of samples of each group that were classified correctly, we notice that for many tasks, the tools provided very different recall rates based on diagnostic (e.g., in Cartoons, ASD: $0.77$, TD: $0.18$). 
We observe large variability in performance for labeling the ASD group when looking at different tasks. When considering F1-scores, a measure used for imbalanced datasets typically found in clinical problem scenarios, OpenSMILE generally performed worse than Praat. The key observation we make here is that classification performance can vary widely depending on task context, tool used, and across each group, highlighting the importance of more nuanced reporting of such classification performance and data pre-processing instead of relying on a single metric.


\begin{table*}[h]
    \centering
    \caption{Two-way ANOVA results to test whether the five speech features differ based on diagnostic group, gender group, and the tool used to extract them. Partial eta squared ($\eta^2_p$) is reported to show effect size alongside $p$-value. In the presence of a significant interaction between tool and group, post-hoc tests were performed to adjust the results for each tool (\textbf{P}raat and \textbf{O}penSMILE)}.
    \begin{tabular}{lccccc}
        \toprule
        \textbf{Task} & \textbf{Feature} & \textbf{Gender Group ($p$ / $\eta^2_p$)} & \textbf{Tool ($p$ / $\eta^2_p$)} & \textbf{Interaction ($p$ / $\eta^2_p$)} & \textbf{ Post-hoc (P $\vert$ O)} \\
        \midrule
        
        Description & Mean Pitch  & $<0.05~/~0.093 $  & $<0.001~/~0.973 $  & $<0.05~/~0.071$ & $<0.05$ $\vert $ ~~~~~~~~~~~\\
        Demonstration & Mean Pitch  & $<0.05~/~0.103$  & $<0.001~/~0.933$  & $<0.05~/~0.073$ & $<0.01$ $\vert$ $<0.05$\\
        Loneliness & Pitch Range & $<0.05~/~0.113$  & $<0.001~/~0.820$  & $<0.05~/~0.098$  & $<0.05$ $\vert$ ~~~~~~~~~~~ \\
        Loneliness & Pitch Std & $<0.05~/~0.110$  & $<0.001~/~0.788$  & $<0.05~/~0.106$ &  $<0.05$ $\vert$ ~~~~~~~~~~~ \\
        Creating & Speech Rate & $<0.05~/~0.084$ & $<0.001~/~0.994$ & $<0.05~/~0.101$ &  $<0.05$ $\vert$ ~~~~~~~~~~~\\
        Loneliness & Speech Rate & $<0.05~/~0.102$  & $<0.001~/~0.985$ & $<0.05~/~0.105$ &  $<0.05$ $\vert$ ~~~~~~~~~~~\\
        \hline
        \hline
        \textbf{Task} & \textbf{Feature} & \textbf{Diagnostic Group ($p$ / $\eta^2_p$)} & \textbf{Tool ($p$ / $\eta^2_p$)} & \textbf{Interaction ($p$ / $\eta^2_p$)} & \textbf{ Post-hoc (P $\vert$ O)}\\
        \hline
        Cartoons & Speech Rate  & $<0.01~/~0.166$  & $<0.001~/~0.987$ & $<0.05~/~0.078$ &  $<0.05$ $\vert$ ~~~~~~~~~~~\\
        Construction & Pitch Std & $<0.05~/~0.102$  & $<0.001~/~0.946$ & $<0.05~/~0.105$ &  $<0.10$ $\vert$ ~~~~~~~~~~~\\
        \bottomrule
    \end{tabular}
    \label{tab:anova_combined_results}
\end{table*}

\begin{table*}[th]
  \caption{ASD diagnosis classification performance using a random forest classifier for each task, comparing Praat and OpenSMILE feature sets extracted from participant data. The results are reported separately for ASD and Typically Developed (TD) groups. 
  }
  \label{tab:classification_tasks_child}
  \centering
  \begin{tabular}{ l c c c c c c }
    \toprule
    \textbf{Task} & 
    \multicolumn{2}{c}{\textbf{Precision (Praat | OpenSMILE)}} & 
    \multicolumn{2}{c}{\textbf{Recall (Praat | OpenSMILE)}} & 
    \multicolumn{2}{c}{\textbf{F1-score (Praat | OpenSMILE)}} \\
    \cmidrule(lr){2-3} \cmidrule(lr){4-5} \cmidrule(lr){6-7}
    & \textbf{ASD} & \textbf{TD} & \textbf{ASD} & \textbf{TD} & \textbf{ASD} & \textbf{TD} \\
    \midrule
    Friends      & 1.00 | 1.00 & 1.00 | 1.00 & 0.37 | 0.30 & 0.59 | 0.53 & 0.51 | 0.45 & 0.73 | 0.67 \\
    Creating     & 1.00 | 1.00 & 1.00 | 1.00 & 0.41 | 0.43 & 0.53 | 0.60 & 0.56 | 0.57 & 0.69 | 0.74 \\
    Cartoons     & \cellcolor{mycolor}1.00 | 0.87 & \cellcolor{mycolor}1.00 | 0.73 & 0.77 | 0.80 & \cellcolor{mycolor}0.18 | 0.23 & 0.86 | 0.88 & \cellcolor{mycolor} 0.28 | 0.33 \\
    Break        & 1.00 | 1.00 & 1.00 | 0.93 & 0.68 | 0.65 & 0.28 | 0.27 & 0.81 | 0.78 & \cellcolor{mycolor} 0.42 | 0.41 \\
    Demonstration & \cellcolor{mycolor}0.77 | 0.85 & 0.93 | 1.00 & \cellcolor{mycolor}0.27 | 0.35 & 0.51 | 0.58 & \cellcolor{mycolor} 0.38 | 0.47 & 0.63 | 0.71 \\
    Telling      & 1.00 | 1.00 & 0.93 | 1.00 & 0.63 | 0.66 & 0.26 | 0.34 & 0.77 | 0.79 & 0.39 | 0.49 \\
    Loneliness   & 1.00 | 0.92 & \cellcolor{mycolor}1.00 | 0.85 & 0.59 | 0.57 & \cellcolor{mycolor}0.52 | 0.38 & 0.72 | 0.70 & 0.64 | 0.50 \\
    Description  & 1.00 | 1.00 & 1.00 | 1.00 & 0.47 | 0.52 & 0.48 | 0.53 & 0.63 | 0.66 & 0.63 | 0.67 \\
    Conversation & 0.92 | 1.00 & 1.00 | 1.00 & \cellcolor{mycolor}0.53 | 0.67 & 0.34 | 0.35 & \cellcolor{mycolor} 0.66 | 0.79 & 0.49 | 0.50 \\
    Make         & 1.00 | 1.00 & 1.00 | 0.92 & 0.50 | 0.52 & 0.41 | 0.41 & 0.64 | 0.67 & 0.56 | 0.53 \\
    Emotions     & 1.00 | 1.00 & 1.00 | 1.00 & 0.64 | 0.61 & 0.43 | 0.36 & 0.76 | 0.74 & \cellcolor{mycolor} 0.57 | 0.50 \\
    Social       & \cellcolor{mycolor}1.00 | 0.85 & 1.00 | 1.00 & 0.39 | 0.37 & 0.60 | 0.65 & 0.54 | 0.47 & 0.73 | 0.77 \\
    Construction & 1.00 | 1.00 & 1.00 | 1.00 & 0.72 | 0.68 & \cellcolor{mycolor}0.21 | 0.36 & 0.83 | 0.79 & \cellcolor{mycolor} 0.34 | 0.50 \\
    Joint        & 0.89 | 0.89 & 1.00 | 1.00 & 0.25 | 0.34 & 0.63 | 0.67 & 0.36 | 0.44 & 0.75 | 0.78 \\
    \bottomrule
  \end{tabular}
  
\end{table*}

\section{Discussion}

The current project is a first step toward better understanding the measures provided by open-source speech analysis tools and how they are affecting our understanding of mental health disorders such as Autism. Researchers have delved into the use of such tools with sometimes minimal understanding. For example, the two tools described have default settings or give the opportunity to set one when needed. However, many papers do not report whether the default parameters were employed or how the settings were chosen. To our knowledge, we are the first to compare feature outputs between the two tools using the default parameters. As shown, they produce significantly different values for the same audio input, raising concern over the comparability of results across studies. Therefore, as a first step, we recommend that future projects report not only the tool used but the parameters set (and the reasons for those choices) when interested in capturing behavioral data, as has been encouraged by Interspeech this year. 

Second, if the intent is to train a classification model, the importance of feature extraction becomes even more important. If the ultimate goal of this field of research is to eventually lessen the cognitive load of trained professionals and aid them in their diagnostic and treatment process, then we need to establish the validity of the measures being used in classification models.

The current project does not claim that one tool or feature is superior to another. \blue{However, we show that widely used tools can yield significantly different results from the same data, affecting the interpretation of behavioral features}. Our findings here point to enough discrepancies to give us pause. Should we trust one model over another when we are unaware as to how the features were measured? How do these decisions influence the performance of the machine learning models we publish about? 
To inform clinical applications, we need to first establish the clinical validity of such measures. 
Just because extracted features perform well on classification models does not mean they will provide new clinical insights or help.



\subsection{Limitations}

Experimental results presented here used utterance-level audio files created using automatic speaker diarization. Results may improve with manual diarization, but automatic utterance processing is more feasible in practical setting. 
 The language of this dataset is English only, and whether our finding generalizes to other languages is left for the future. \blue {We also acknowledge the gender imbalance in the dataset used in our work. Balanced datasets are difficult to obtain for such domains~\cite{nordahl2023we}, and our study highlights the broader need for large, gender-balanced data for ML-based speech studies, including autism research.} 
Here, we focused on OpenSMILE and Praat to compare speech features. Future work can expand the comparison to other similar tools such as librosa~\cite{mcfee2015librosa} and Prosodypro~\cite{xu2013prosodypro}.

\section{Future Work}
We plan to first deal with the limitations of our approach here and delve into more careful and informed setting of parameters when using open-source tools. Next, we acknowledge that conversation is an interactive activity, and plan to also analyze the behaviors of the psychologist administering the diagnostic, as previous research has found interesting clinically relevant information can be gained from this step~\cite{chowdhury23_interspeech, fusaroli_hearing_2019, lahiri2022interpersonal}. Finally, we would like to explore different ways to capture the rich, time-evolving nature of verbal interaction by moving away from calculating mean features throughout a whole task. Instead, we would like to analyze the possible time-series data set that can be obtained from these tools to observe the change in speech features over the course of a task.



\bibliographystyle{IEEEtran}
\bibliography{references}

\end{document}